\newcommand{\CLASSINPUTtoptextmargin}{2.4cm}
\newcommand{\CLASSINPUTbottomtextmargin}{4.6cm}
\documentclass[conference]{IEEEtran}
\IEEEoverridecommandlockouts
\usepackage{cite}
\usepackage{amsmath,amssymb,amsfonts}
\usepackage{algorithmic}
\usepackage{graphicx}
\usepackage{textcomp}
\usepackage{xcolor}
\usepackage{url}
\def\BibTeX{{\rm B\kern-.05em{\sc i\kern-.025em b}\kern-.08em
    T\kern-.1667em\lower.7ex\hbox{E}\kern-.125emX}}
\def\BibTeX{{\rm B\kern-.05em{\sc i\kern-.025em b}\kern-.08em T\kern-.1667em\lower.7ex\hbox{E}\kern-.125emX}}
\begin{document}

\title{IoT Botnet Detection Using an Economic Deep Learning Model\\
}

\author{\IEEEauthorblockN{Nelly Elsayed}
	\IEEEauthorblockA{\textit{School of Information Technology} \\
		\textit{Univesity of Cincinnati}\\
		Ohio, United States\\
		elsayeny@ucmail.uc.edu}
	\and
	\IEEEauthorblockN{Zag ElSayed}
	\IEEEauthorblockA{\textit{School of Information Technology} \\
	\textit{Univesity of Cincinnati}\\
	Ohio, United States\\
	elsayezs@ucmail.uc.edu}
	\and
	\IEEEauthorblockN{Magdy Bayoumi}
	\IEEEauthorblockA{\textit{Dept. of Electrical and Computer Engineering}\\
		\textit{University of Louisiana at Lafayette}\\
	Louisiana, United States\\
	magdy.bayoumi@louisiana.edu}
}

\thispagestyle{empty}

\begin{huge}
	IEEE Copyright Notice
\end{huge}

\vspace{5mm} 

\begin{large}
	Copyright (c) 2023 IEEE
\end{large}

\vspace{5mm} 

\begin{large}
	Personal use of this material is permitted. Permission from IEEE must be obtained for all other uses, in any current or future media, including reprinting/republishing this material for advertising or promotional purposes, creating new collective works, for resale or redistribution to servers or lists, or reuse of any copyrighted component of this work in other works.
\end{large}

\vspace{5mm} 

\begin{large}
	\textbf{Accepted to be published in:} IEEE World AI IoT Congress (AIIoT) 2023; June 7 - June 10 , 2023.
	https://https://worldaiiotcongress.org/
	
\end{large}

\vspace{5mm} 

\maketitle

\begin{abstract}
	The rapid progress in technology innovation usage and distribution has increased in the last decade. The rapid growth of the Internet of Things (IoT) systems worldwide has increased network security challenges created by malicious third parties. Thus, reliable intrusion detection and network forensics systems that consider security concerns and IoT systems limitations are essential to protect such systems. IoT botnet attacks are one of the significant threats to enterprises and individuals. Thus, this paper proposed an economic deep learning-based model for detecting IoT botnet attacks along with different types of attacks. The proposed model achieved higher accuracy than the state-of-the-art detection models using a smaller implementation budget and accelerating the training and detecting processes.
\end{abstract}


\section{Introduction}

The Internet of Things (IoT) is the network between physical devices and cyberspace, such as smart ring bells, smart light bulbs, surveillance cameras, thermostats, smart medical devices, and traffic control systems. It is the automation of all kinds of devices. IoT is also considered a communication system that is interconnected via wireless or wired communication technologies~\cite{hussain2020machine}. The main concept of the Internet of things first appeared in 1982 through the Carnegie Mellon University project that modified a Coca-Cola vending machine to be able to report the machine inventory and whether newly loaded drinks were cold or not~\cite{machine2014only}. This idea became the first ARPANET-connected appliance~\cite{machine2014only}. Then, in 1999, Kevin Ashton from Procter and Gamble Cooperation (P\&G) coined the term "Internet of Things" to describe the system where the physical world connected to the Internet via ubiquitous sensors~\cite{ashton2009internet}.

\begin{figure}[htbp]
	\centerline{\includegraphics[width=7cm,height=6cm]{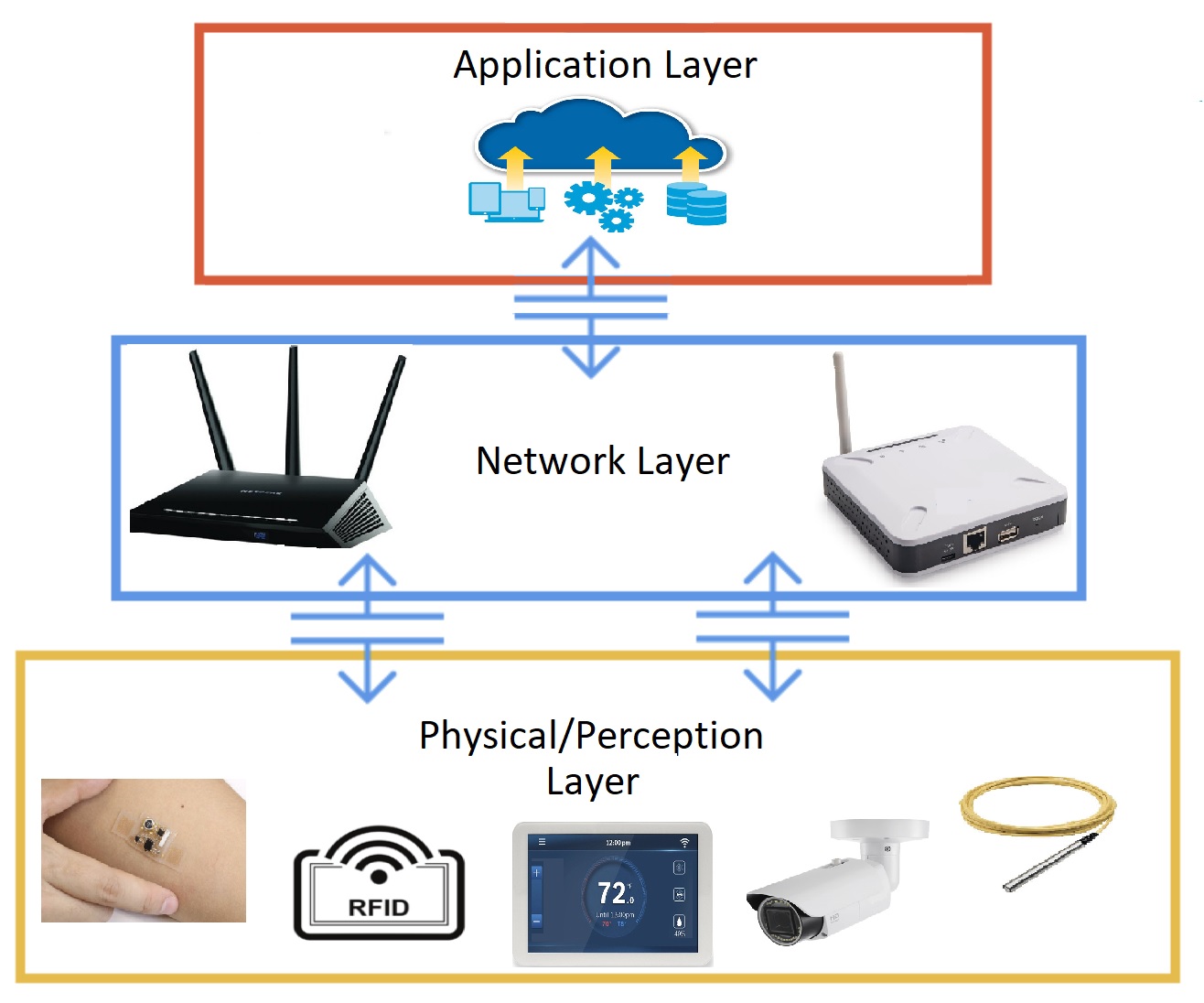}}
	\caption{The three layered architecture framework of the IoT security~\cite{azumah2021deep}.}
	\label{IoT_architecture}
\end{figure}

\begin{table*}[t!]
	\caption{IoT botnets and their attack types.}
	\begin{center}
		
		\resizebox{\textwidth}{!}{	
			\begin{tabular}{|l|c|c|c|c|}
				
				\hline
				\textbf{IoT Botnet}&\textbf{Attack Types}& \textbf{Exploit Types} & \textbf{Open} & \textbf{First}\\
				\textbf{}&\textbf{}& \textbf{} & \textbf{Source} & \textbf{Occured}\\
				\hline
				Chuck Norris~\cite{ChuckNorris} & DDoS& SSH, Telnet& No & 2009\\
				PsybOt~\cite{vdurfina2013psybot} & DDoS&SSH, Telnet& No& 2009\\
				Tsunami/Kaiten~\cite{kaiten,kambourakis2017mirai}& DDoD&SSH, Telnet&  No &2010\\
				Aidra~\cite{vignau201910} & DDoD& Telnet & Yes & 2012\\
				Carana~\cite{malecot2013carna}& N/A & Telnet & No & 2013 \\
				Bashlite/Gafgyt/Qbot~\cite{mcnulty2022iot} &DDoS &Telnet&Yes &2014\\
				LightAidra~\cite{coltellese2019triage}& DDoS& Telnet& Yes&2014\\
				muBoT~\cite{arivudainambi2019malware} & DDoD& SSH, Injection & Yes&2014\\
				Wifach~\cite{ferronato2020iot} & Secure endpoint& Telnet &Yes&2014\\
				XOR DDoS~\cite{bederna2020cyber}&DDoS & SSH& No & 2014\\
				Amnesia~\cite{prokofiev2018method}&DDoS& RCE &No&2016\\
				Bashlite II~\cite{spring2016bashlite}&DDoS&SSH, Telnet&No&2016\\
				BillGates botnet~\cite{dehoneynet}&DDoS&SSH&No&2016\\
				Hajime~\cite{edwards2016hajime}&Secure endpoint& Telnet, TR-069& No& 2016\\
				IRC Telnet~\cite{hamza2020iot}&DDoS& Telnet&No&2016\\
				Lua Bot~\cite{geenens2019iot}&DDoS&N/A&No&2016\\
				NyaDrop~\cite{matthews2021analysis}&N/A&Telnet&No&2016\\
				Mirai~\cite{kambourakis2017mirai}&DDoS&Telnet&Yes&2016\\
				Remaiten~\cite{paganini2016linux,cozzi2020tangled}&DDoS&Telnet&No&2016\\
				TheMoon~\cite{TheMoon}&Spying, MPAB-for-hire, PDoS&4 CVs&No&2016\\
				BrickerBot~\cite{cimpanu2017brickerbot,shobana2018iot}&DDoS, PDoS endpoint&Telnet&No&2017\\
				Persirai~\cite{yeh2017persirai}&DDoS&UPnP BF&No&2017\\
				Reaper~\cite{torabi2018inferring,berasaluce2019cybercrime}&DoS& HTTP& No& 2017\\
				Satori~\cite{geenens2019iot,satori}&DDoS& Brutforce&Yes&2017\\
				DoubleDoor~\cite{backdoorbotNet}& N/A &password  & No &2018\\
				Hide and seek~\cite{hideAndSeek,csendroiu2018hide}&  Spying& Telnet, HTTP &No  &2018\\
				JenX~\cite{jenx}& DDoS, DDo-for-hire & TCP & Yes&2018\\
				Masuta~\cite{pereira2009machine}& BF &  DDoS, DDoS-for-hire& Yes &2018\\
				Omni~\cite{farooq2019modeling}& Cryptojacking & N/A & No &2018\\
				Owari~\cite{owari}& MPAB-for-hire & password & Yes &2018\\
				Sora~\cite{sora}&DDoS, MPAB-for-hire& Brutforce &No&2018\\
				VPNFilter~\cite{VPNFilter}&Apying, PDoS&14 CVEs&No&2018\\
				Wicked~\cite{mcdermott2019evaluating}&DDoS&N/A&No&2018\\
				Bashlite III~\cite{javed2019multi}&DDoS, Cryptojacking& Telnet, UPnP&No& 2019\\
				Echobot~\cite{Echobot}&DDoS&+50 RCE exploits&No&2019\\
				Stealthworker~\cite{Stealthworker}&Ransomeware&SSH, Brutforce&No&2019\\
				Dark Nexus~\cite{DarkNexus}&DDoS&Telnet&No&2020\\
				Fronton~\cite{Fronton}&DDoS, Dosinformation&N/A&No&2020\\
				Kaiji~\cite{Kaiji}&DDoS&SSH&No&2020\\
				Mozi~\cite{Mozi}&DDoS, Data Exfiltration, Payload execution&Telnet&No&2020\\
				Ttint~\cite{Ttint}&DDoS&zero-day&No&2020\\
				Dark.IoT~\cite{darkIoT}&DDoS& zero-day, CVE &No&2021\\
				Meris~\cite{meris,najar2022ddos}&DDoS &RouterOS&No&2021\\
				EnemyBot~\cite{EnemyBot}&DDoS& SSH, Telnet, Brutforce, RCE&Yes&2022\\			
				Zerobot~\cite{zeroBot1,zeroBot2,zeroBot3}&DDoS&SSH, Telnet&No&2022\\
				\hline
			\end{tabular}
		}
		\label{IoTBots}
	\end{center}
\end{table*}
Regardless of the simple design of the IoT device, it contains a complex architecture, connection layers, and network components that transfer data and requires a robust security architecture design~\cite{hassija2019survey}. As the number of different types of applications and cloud-based systems involve IoT, the security threats on IoT inherently grow. Further, one of the most critical issues of the IoT systems has become security due to IoT hardware and software resource limitations~\cite{sandell2018application}. Maintaining the security of IoT devices is crucial since IoT devices may contain sensitive data such as a user's personal information, location, and activity.

\begin{figure*}[htbp]
	\centerline{\includegraphics[width=12cm,height=7.5cm]{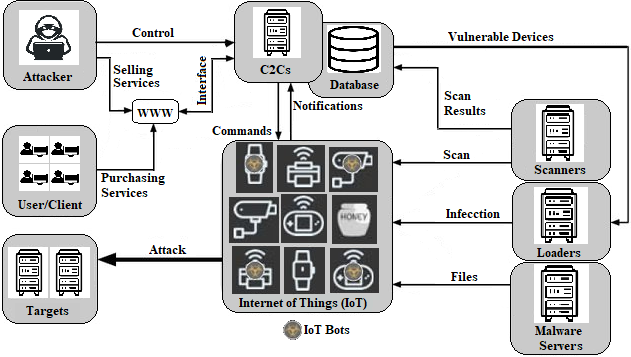}}
	\caption{A diagram of how the IoT botnet attack is performed.}
	\label{IoTBotnet_Overview}
\end{figure*}
Distributed Denial of Service Attack (DDoS) is one of the most common attacks on IoT devices~\cite{abomhara2015cyber,meneghello2019iot,kolias2017ddos}. Such attacks are primarily performed using different IoT botnets, a group of computers that have been infected by malware and come under the control of a malicious actor. Botnets can be designed to accomplish illegal or malicious tasks, including sending spam, stealing data, ransomware, fraudulently clicking on ads, or distributed denial-of-service (DDoS) attacks. Mirai, Omni, EmemyBot, Persirai, and Hide'n Seek are examples of such botnets~\cite{kolias2017ddos,mcnulty2022iot}. Network security monitoring and IoT botnet detection systems are crucial to prevent such malicious attacks~\cite{zaghloul2021green,soltan2018blackiot,elsayed2021autonomous}. Thus, in this paper, we address the IoT botnet detection systems and target the green AI implementation concept where the model implementation reduction maintaining the performance is one of the primary goals to reduce power usage~\cite{schwartz2020green}. 

In this paper, we proposed an ecomonic deep learning based model for IoT botnet detection. The proposed model achieves a significant accuracy to detect IoT botnets using a significant smaller implementation budget compared to the state-of-the-art models. The contribution of this paper can be summarized as follows:
\begin{enumerate}
	\item We proposed an IoT botnet detection system that can classify the botnet attack category and sub category using deep learning. 
	\item The proposed model reduce the implemenration budget by approximate 76\% compared to the current state-of-the-art models in addition to increasing the detection performance. 
	By reducing the implementation budget, the proposed model requires less training time and power to implement the proposed model, leading to green AI model implementation.
\end{enumerate}

This paper's structure is as follows: Section~\ref{IoT_layers} describes the IoT architecture components. Section~\ref{IoT_botnets} describes the IoT botnets and the IoT botnet attacks threats. Section\ref{proposed_model} describes the proposed model components and architecture. Finally, Section~\ref{results_section} provides the experimental results and analysis of the proposed IoT botnet detection model.
\section{IoT Security Layers}\label{IoT_layers}

The IoT security layered architecture consists of three layers: the physical/perception layer, the network layer, and the application layer. Figure~\ref{IoT_architecture} demonstrated the IoT security architecture layers. The perception layer contains the physical component of the IoT environment, including the sensors and different sensing devices. The network layer consists of the network devices, servers, and the smart things of the IoT. The application layer is responsible for providing the user with specific services via different applications~\cite{sethi2017internet}. 

\subsection{Botnets}\label{IoT_botnets}
The IoT botnet consists of two primary components and an additional (optional) component. The first primary component is the  Bot itself which can be an agent or an end zombie IoT device that performs the DDoS attacks on a command~\cite{angrishi2017turning}. The second primary component is the command-and-control servers (C2Cs) that are used to control the bots. The four optional components are the scanners, the reporting server, the loaders, and the malware distribution server. The scanners are used for scanning vulnerable IoT devices. The reporting server scan reports or collects the results or scan reports from external scammers or IoT bots. The loaders first log on to the vulnerable IoT device and then start to instruct the vulnerable IoT device to download malware. The malware distribution server is the location where the malware code is stored so that the loader navigates the infected IoT device to download it~\cite{angrishi2017turning}. Figure~\ref{IoTBotnet_Overview} shows a diagram of how the IoT botnet attack is performed~\cite{marzano2018evolution,al2020lightgbm}. 

There are several IoT botnet attacks that cause different types of cyber attacks, exploits, or both. Table~\ref{IoTBots} describes the IoT botnets from 2009 to 2023, including their attack types, the type of exploit they are performed, the year of the first infection, the source of infection, and whether it is open-source or not. The IoT botnets are ordered in the table from the oldest to the newest, followed by an alphabetical order for the botnets that have the same year of occurrence. 

\begin{figure*}[t]
	\centerline{\includegraphics[width=11cm,height=10.5cm]{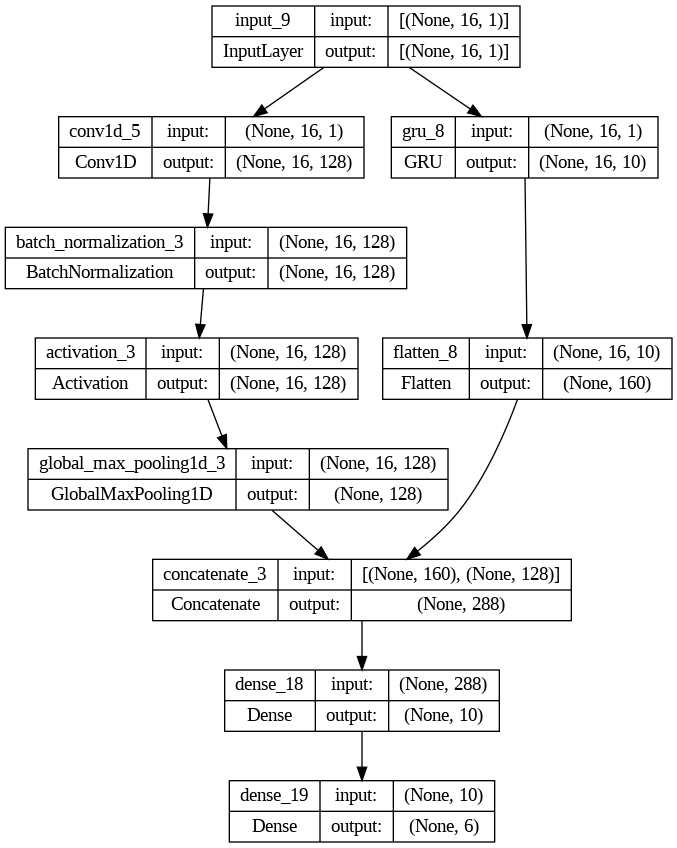}}
	\caption{The proposed model architecture diagram.}
	\label{model_architecture}
\end{figure*}

\begin{table}[t]
	\caption{The proposed model summary.}
	\begin{center}
			\small
			\begin{tabular}{|l|l|l|}
				\hline
				\textbf{Layer (type) }   &               \textbf{Output Shape}    &     \textbf{Param \#}        \\            
				\hline
				InputLayer   &       [(None, 16, 1)]     &   0        \\                            
				
				Conv1D           &        (None, 16, 128)   &      512    \\              
				
				BatchNormalization &  (None, 16, 128)     &   512        \\

				GRU          &         (None, 16, 10)      & 390         \\         
				
				Activation &      (None, 16, 128)   &   0        \\
				
				Flatten    &         (None, 160)    &      0     \\                 
				
				GlobalMaxPooling1D&  (None, 128)    &     0         \\

				Concatenate   &    (None, 288)       &    0          \\

				dense (Dense)        &          (None, 10)    &       2890          \\    
				
				dense\_1 (Dense)        &        (None, 6)      &      66        \\                
				
				\hline
				
			\end{tabular}
		\label{model_summary}
	\end{center}
\end{table}

\section{Proposed Model}\label{proposed_model}

\begin{figure}[t]
	\centering
	\includegraphics[width=7cm,height=5cm]{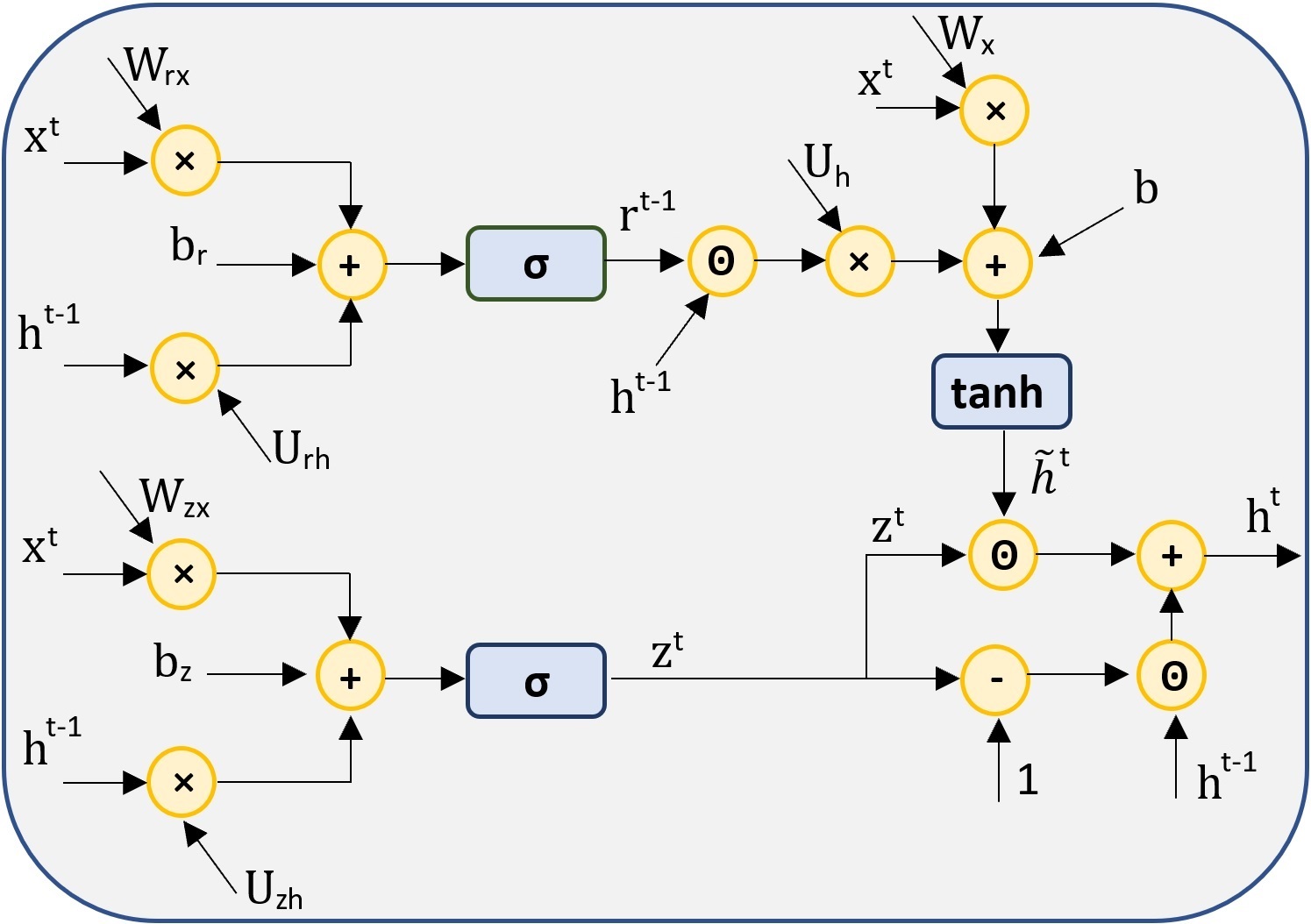}
	\caption{The GRU architecture diagram showing the corresponding weights of each component.}
	\label{gru_weights}
\end{figure}

In this paper, we proposed an economic system to detect IoT botnets using a deep learning model. Our proposed model reduces the implementation budget by approximately 79\% compared to the state-of-the-art deep learning-based models for IoT botnet detection. Our proposed model used efficient, low-cost implementation deep learning architecture, which is able to achieve efficiency and outperform the state-of-the-art. Therefore, the proposed model requires less implementation power leading to green AI system implementation.

The proposed model architecture is shown in Figure.~\ref{model_architecture}. The proposed model concatenates the gated recurrent neural network capable of learning the temporal information in the data and the spatial information captured via the convolution neural network. Selecting the GRU over other recurrent architecture based on the findings in~\cite{elsayed2018deep} where combining the GRU with the CNN has higher accuracy results compared to combining the LSTM with CNN. Moreover, the GRU requires less time and implementation budget from software and hardware aspects as it has a smaller architecture than the LSTM~\cite{chung2014empirical,zaghloul2021fpga}. The GRU consist only of two gates: reset and update gates. At time $t$, the current output of the GRU, $h^{(t)}$ is calculated as follows: 
\begin{flalign}
	&z^{(t)} = \sigma (W_{zx} x^{(t)} + U_{zh} h^{(t-1)} + b_z) \label{eqn:z_gate}\\
	&r^{(t)} = \sigma (W_{rx} x^{(t)} + U_{rh} h^{(t-1)}+ b_r)\label{eqn:r_gate}\\
	&{\tilde h}^{(t)} = \mathrm{tanh} (W_{x} x^{(t)} + U_{h}( r^{(t)} \odot h^{(t-1)})+b)\label{eqn:h_hat}\\
	&h^{(t)} = (1-z^{(t)})\odot h^{(t-1)}+z^{(t)}\odot{\tilde h}^{(t)}\label{eqn:h}
\end{flalign}
\noindent
where $x^{(t)}$ is the input at time $t$ and $h^{(t-1)}$ is the reccurrent output at time $t-1$. $z^{(t)}$ and $r^{(t)}$ are the GRU update and reset gates. ${\tilde h}^{(t)}$ is the output candidate activation. The feedforward weights of the update gate, reset gate, and output candidate activation are $W_{zx}$, $W_{rx}$, and $W_{x}$, respectively. The recurrent weights of the update gate, reset gate, and output candidate activation are $U_{hz}$, $U_{hr}$, and $U_{h}$, respectively. The biases of the update gate, reset gate and the output candidate activation are $b_z$, $b_r$ and $b$, respectively.
The GRU architecture with weights and biases made explicit is shown in Figure~\ref {gru_weights}. $\sigma$ is the sigmoid function (Eqn.~\ref{sigmoid}) and $tanh$ is the hyberbolic tangent (Eqn.~\ref{tanh})~\cite{elsayed2018empirical}. 

\begin{equation}\label{sigmoid}
	f(x)= \sigma(x) = \frac{e^x}{e^x+1} = \frac{1}{1+e^{-x}} ,
\end{equation}
\begin{equation}\label{tanh}
	\tanh(x)=  \frac{\sinh(x)}{\cosh(x)}= \frac{e^{x}-e^{-x}}{e^{x}+e^{-x}} = \frac{e^{2x}-1}{e^{2x}+1} 
\end{equation}

The convolutional layers within the CNN learn to extract the feature representations from the data without the requirement of data preprocessing stages. The global average pooling layer is a pooling operation that designed to replace fully connected layers in  the convolutional neural networks~\cite{lin2013network}. We used the global average pooling layer~\cite{boureau2010theoretical} to interpret the classes and to reduce the number of trainable parameters compared to the fully connected layer without accuracy sacrification. The concatenation layer concatenates the input from the GRU and the convolutional block. Thus, the total number of parameters of the proposed model is 4370, where 4114 parameters are trainable, and 256 are non-trainable parameters. Thus, the model achieves high detection accuracy using a low implementation budget compared to the current state-of-the-art models. The proposed model summary is shown in Table~\ref{model_summary}, where each layer of the proposed model, its corresponding output shape, and the number of trainable parameters are provided.    

\begin{table}[htbp]
	\caption{The proposed model overall results and statistics.}
	\begin{center}
			\small
			\begin{tabular}{|l|c|}
				\hline
				\textbf{Metrics}&\textbf{Value}\\
				\hline
				Train accuracy& 99.73\%\\
				Test Accuracy& 99.25\%\\
				95\% CI&(0.99239,0.99279) \\
				F1-Score&0.9925\\
				Kappa & 0.98307 \\
				Hamming Loss&0.00741\\
				RCI &0.9828 \\
				\#Parameters&4370\\
				Train time& 25.21 (minutes)\\
				\hline
			\end{tabular}
		\label{model_results}
	\end{center}
\end{table}	

\begin{table*}[t]
	\caption{The proposed model testing stage classes statistics.}
	\begin{center}
			\small
			\begin{tabular}{|l|c|c|c|c|c|c|}
				\hline
				\textbf{}&\textbf{Class 0}&\textbf{Class 1}&\textbf{Class 2}&\textbf{Class 3}&\textbf{Class 4}&\textbf{Class 5}\\
				\hline
				ACC(Accuracy)                                  &                   0.99993     &  0.99984    &   0.99287     &  0.9928    &    0.99989    &   0.99985    \\   
				AGF(Adjusted F-score)                        &                     0.99997     &  0.99983    &   0.68433    &   0.86309    &   0.94874    &   0.0       \\    
				AGM(Adjusted geometric mean)                 &                     0.99991     &  0.99985    &   0.85544     &  0.92441   &    0.97189   &    0        \\     
				AUC(Area under the ROC curve)                  &                   0.99993      & 0.99984     &  0.75543    &   0.86312  &     0.94542   &    0.5 \\          
				AUCI(AUC value interpretation)               &                     Excellent   &  Excellent   &  Good      &    Very Good    & Excellent   &  Poor    \\
				ERR(Error rate)                               &                    7e-05       &  0.00016 &      0.00713    &   0.0072     &   0.00011   &    0.00015  \\
				F1-Score &         0.99994    &   0.99982  &     0.41423    &   0.77933   &    0.91446   &    0.0 \\
				Precision & 0.99989    &   0.99982    &   0.34615    &   0.83744    &  0.93933    &   None    \\
				False Negative      &4     &        60       &     1734     &     3463    &      55       &     107 \\
				False Positive & 4      &        57         &     3487       &     1806         &   29      &        0      \\
				True Positive &  396568    &    318277     &   1846       &   9304        &  449    &       0    \\
				True Negative & 335251     &   413473    &    724800     &   717294     &   731334  &      731760    \\
				Y(Youden index)                   &                                0.99986    &   0.99967     &  0.51085    &   0.72624     &  0.89083   &    0.0    \\       
				dInd(Distance index)               &                               0.00013   &    0.00023    &   0.48438    &   0.27126    &   0.10913    &   1.0      \\     
				sInd(Similarity index)              &                              0.99991     &  0.99983    &   0.65749   &    0.80819     &  0.92284     &  0.29289  \\    
				\hline
			\end{tabular}
		\label{class_statistics}
	\end{center}
\end{table*}

\section{Experiments and Results}\label{results_section}

\subsection{Dataset}
For our model design and testing, we used the UNSW 2018 IoT Botnet Dataset benchmark~\cite{koroniotis2019towards}. This data has been created using a realistic network environment in the Cyber Range Lab of UNSW Canberra. Thus, this dataset is sufficient to create IoT botnet detection models than the non-realistic network environment datasets. The dataset has been collected using a weather station, smart fridge, motion-activated lights, remote-activated garage door, and smart thermostat. The dataset has a binary categorization indicated by $0$ as normal and $1$ as an IoT botnet attack. In addition, It has subcategories that explicitly indicate the different types of attacks leading to six different subcategories where Class $0$ indicates normal, Class (1) indicates DDoS TCP attack, Class (2) indicates DDoS UDP attack, Class (3) indicates DoS HTTP attack, Class (4) indicates OS Fingr/t attack, and Class (5) indicates Data ex-filtration attack. The dataset's full description and details are provided in~\cite{koroniotis2019towards}.

\begin{figure}[htbp]
	\centerline{\includegraphics[width=7cm,height=4cm]{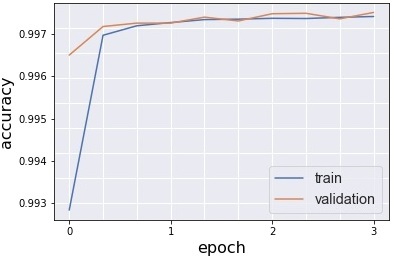}}
	\caption{The training versus validation accuracy over the epochs.}
	\label{accuracy}
\end{figure}

\begin{figure}[htbp]
	\centerline{\includegraphics[width=7cm,height=4cm]{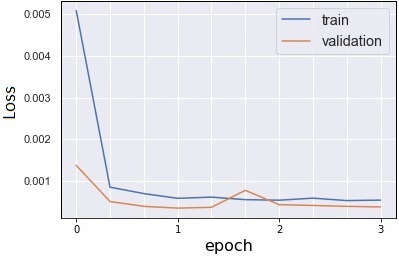}}
	\caption{The training versus validation loss over the epochs.}
	\label{loss}
\end{figure}

\subsection{Experiment Parameters}

The proposed IoT botnet detection model was implemented on Google Colab using Python 3.8, Tensorflow 2.9.2, Keras 2.9.0, and NumPy 1.21.6. In order to set a fair comparison between our model and the state-of-the-art models for IoT botnet detection~\cite{koroniotis2019towards}. The model was trained for four epochs, and the batch size was set to 10. The UNSW 2018 IoT Botnet Dataset benchmark has already been split into training, and testing datasets with sizes 2927524 and 731867 attributes, respectively. We split the training dataset to train and validation datasets with ratios of 90\% and 10\%, respectively. For the GRU, the number of unrollments was set to 10, and the weights were initialized using a Truncated Normal initializer. The 1D convolutional layer kernel size was set to three, and the weights were initialized using the he uniform initializer. We used the RMSProp optimization function to compile the model, and the loss was calculated using the categorical cross-entropy function. Figure~\ref{accuracy} shows the training versus validation training accuracy over the epochs. Figure~\ref{loss} shows the training versus validation training loss over the epochs.

Table~\ref{model_results} shows the proposed model's overall results and statistics including the training and testing accuracies, accuracy macro, precision, recall, F1-Score, specificity. The F1-Score calculated by:
\begin{flalign}
	precision &= \frac{True Positive}{True Positive+False Positive}\\
	recall&= \frac{True Positive}{True Positive+False Negative}\\
	\mathit{f1\textnormal{-}score} &= 2 \times \frac{precision \times recall}{precision + recall}
\end{flalign}
\noindent
where TP, FP, FN stand for true-positive, false-positive and false-negative, respectively. The kappa value~\cite{czodrowski2014count}, and the number of trainable parameters on the proposed model are shown in Table~\ref{model_results}. The kappa value is calulated using the formula:
\begin{equation}
	\kappa = \frac{p_o - pe}{1-p_e} = 1- \frac{1-p_o}{1-p_e}
\end{equation}
\noindent
where $p_o$ is the observed positive recognition, and $p_e$ is the expected positive recognition. Hamming loss defines the fraction of wrong labels to the total number of labels. In multi-class classification, the hamming loss is calculated using the hamming distance between the actual and predicted values.

Table~\ref{class_statistics} shows the statistical analysis of the proposed model results over detecting each class (six classes) of the dataset subcategories.

Table~\ref{comparison_table} shows a comparison between the proposed model and the current state-of-the-art architectures for IoT botnet attack detection where N/A indicates that the particular information has not been provided by its authors. For fair gathering of the training time, we re-implement the models that have been clearly described by their authors on the same machine and envrionment that we have implemeted our proposed mode. Our proposed economic model has exceeded the current models' accuracy. In addition, it has a significant reduction in the implementation budget. Moreover, it is faster in both training and testing aspects which is substantial for monitoring the IoT network traffic.

\begin{table*}[htbp]
	\caption{Comparison between the proposed model and the state-fo-the-art training IoT botnet attack detection models.}
	\small
	\begin{center}
			\begin{tabular}{|l|l|c|c|c|c|}
				\hline
				\textbf{Model}&\textbf{Technique}&\textbf{ Dataset}&\textbf{\#Parameters}&\textbf{Accuracy} & \textbf{Train Time (seconds)}\\
				\hline
				Koroniotis et al.~\cite{koroniotis2018towards}&Association Rule Mining&UNSW 2015&N/A&86.45\%&N/A\\
				Koroniotis et al.~\cite{koroniotis2018towards}&Decision Tree&UNSW 2015&N/A&93.23\%&N/A\\
				Koroniotis et al.~\cite{koroniotis2018towards}&Naiive Bayes & UNSW 2015&N/A&72.73\% &N/A\\
				Zeeshan et al.~\cite{zeeshan2021protocol}& LSTM-CNN  & UNSW 2015&N/A&96.32\%&N/A\\
				Koroniotis et al.~\cite{koroniotis2018towards}&ANN &UNSW 2015&N/A&63.97\%& N/A\\
				Koroniotis et al.~\cite{koroniotis2019towards}&SVM-based& UNSW 2018&N/A & 88.372\% & \textbf{1270 }\\
				Koroniotis et al.~\cite{koroniotis2019towards}&RNN-CNN &UNSW 2018& 14316    &99740\%& 8035\\
				Koroniotis et al.~\cite{koroniotis2019towards}&LSTM-CNN &UNSW 2018&14,676  & 99.7419\%& 10482\\
				\textbf{Our model}& GRU-CNN&UNSW 2018&\textbf{4370} & \textbf{99.753\%} & \textbf{1592} \\
				\hline
			\end{tabular}
		\label{comparison_table}
	\end{center}
\end{table*}

\section{Conclusion}
The Internet of Things cybersecurity challenges have increased rapidly due to the rapid growth of IoT systems and the lack of IoT-focused security tools. In this paper, we focused on designing an economic deep learning-based model to detect IoT botnet attacks along with different types of attacks. Our proposed model has exceeded the accuracy of state-of-the-art IoT botnet detection approaches using only 25\% of the implementation budget. Thus, it reduces the implementation budget by approximately 75\%. In addition, it exceeds the state-of-the-art accuracy in detecting the IoT botnet types that help to find a sufficient strategy to suppress the attack. Moreover, it accelerates the model training and testing time. Furthermore, it is simple to implement on both hardware and software. Due to the budget reduction, our proposed model reduces the required power for model implementation leading to a green AI implementation design. 



\bibliography{IoTBot_references}
\bibliographystyle{ieeetr}

\end{document}